# First spectral emissivity study of a solar selective coating in the 150–600 ºC temperature range

I. Setién-Fernández [a], T. Echániz [a], L. González-Fernández [a], R.B. Pérez-Sáez [a,b],, E. Céspedes [c], J.A. Sánchez-García [c], L. Álvarez-Fraga [c], R. Escobar Galindo [c], J.M. Albella [c], C. Prieto [c], M.J. Tello [a,b] a Departamento de Física de la Materia Condensada, Facultad de Ciencia y Tecnología, Universidad del País Vasco, Barrio Sarriena s/n, 48940 Leioa, Bizkaia, Spain

b Instituto de Síntesis y Estudio de Materiales, Universidad del País Vasco, Apdo. 644, 48080 Bilbao, Spain

c Instituto de Ciencia de Materiales de Madrid, (ICMM-CSIC), Campus Cantoblanco, 28049 Madrid, Spain

Abstract

A complete experimental study of temperature dependence of the total spectral emissivity has been performed, for the first time, for absorber–reflector selective coatings used in concentrated solar power (CSP) systems for energy harvesting. The coating consist of double cermet layers of silicon oxide with different amounts of molybdenum over a silver infrared mirror layer. The experimental measurements were carried out by a high accurate radiometer (HAIRL) with controlled atmosphere in the mid-infrared and for temperatures between 150 and 600 ºC. The spectral emissivity is nearly constant in this temperature range. Therefore, the temperature dependence of the total emissivity is given by Planck function. These results were compared with those obtained with the usual calculus using room temperature reflectance spectrum. Finally, the performance of the coating was analyzed by comparison of coated respect to non-coated stainless steel.

1. Introduction

Solar thermal devices are an alternative to produce heat from the sun for heating systems (T <150 ºC) and also to produce solar thermal electricity (150 ºC < T < 800 ºC). In these devices both, the thermal energy storage and the solar thermal collectors (STC) with different configurations (i.e. flat-plate collector and parabolic concentrated collector), have special relevance. In the case of STCs, the solar absorber surface (SAS) is the most important part. A surface that facilitates the conversion of solar radiation into useful heat should possess two important properties: to absorb the incoming solar radiation as much as possible (i.e. high solar absorptivity, α, at the vis–NIR wavelengths) and, at the same time, to retain the collected heat (high thermal reflectivity, R, or low emissivity, ε, at NIR–MIR region [1]).

The most common type of absorber is based on materials which are black in the solar radiation range but transparent for the heat, like metal–ceramic nano-composites ("cermets"). Among all the existing mechanisms, an absorber–reflector tandem consisting of small transition metal particles embedded in a dielectric matrix deposited on a highly infrared reflecting substrate is the most suited method. These thin film coatings offers a high degree of flexibility in order to obtain the desired optical properties to achieve the expected solar selectivity values by changing the thickness, metal volume fraction, and the shape of metal nano particles [2].

Currently, most of the commercial SAS are prepared by magnetron sputtering technology that is a dry, clean and eco-friendly process allowing large area deposition as compared to the electrochemical methods [3–6]. These SASs are composed of two or four homogeneous cermet layers with different metal contents or one cermet layer with a graded refractive index [7,8]. The selectivity can be increased adding more layers [9] but, in this case, the price increases and durability decreases [10], being the double layer cermets the base for the most successful solar selective coatings for medium-high temperature applications [11–14].

Generally the tandem absorbers are degraded at high operating temperatures due to their unstable microstructure, which cause a decrease in the solar selectivity (defined as $\alpha/\varepsilon$). One of the essential requirements of solar selective absorbers is their stability when they operate at high temperatures, from approximately 400 to 600 ºC. Optical properties of these coatings should not deteriorate with the rise of the temperature during the period of use.

To accomplish this, new more efficient selective coatings are needed to get both high solar absorptivity ($\alpha > 0.96$) and low thermal emittance ($\varepsilon < 0.05$) at the working temperature range (400–600 ºC). In fact, for high temperature applications, low $\varepsilon$ is the key parameter, because the thermal radiative losses of the absorbers increase proportionally to $T^4$ [15].

In order to analyze heat losses, a complete knowledge of the radiative properties of the coating structure is essential for their use in high temperature solar collectors. However, a systematic study of direct total spectral emissivity as a function of the temperature for homogeneous cermet of two layers has not been performed yet. For instance, all the measurements reported in the literature were carried out at room temperature or, at most, at 100 ºC. Therefore, the values of emissivity at working temperatures (400–600 ºC) are obtained by extrapolation, which can introduce significant errors in the final result.

This study is focused on the relevance of the high temperature radiometric emissivity techniques in the optical characterization of the SAS. In this paper this experimental technique is applied to study the spectral emissivity behavior of a coating with double layer cermets of silicon oxide with different amounts of metal. The measurements were carried out using a high accurate radiometer with controlled atmosphere in the medium infrared range and for temperatures between 150 and 600 ºC. The results obtained in this study are compared to those obtained with indirect methods.

2. Experimental

2 x 2 cm² plates of stainless steel AISI-321 were used as substrates for the coatings. The substrate roughness was measured using a commercial rugosimeter and the obtained values are showed in Table 1, where $R_a$ is the roughness average, $R_z$ the average maximum height and Rt the maximum height of the profile.

Table 1

Sample surface roughness.

| Ra (µm) | Rz (µm) | Rt (µm) |
|---------|---------|---------|
| 0.13    | 0.87    | 2.14    |

The selective solar coating was prepared by sequential sputtering deposition on steel substrate, air-annealed during 2 h to develop a thermally grown oxide barrier layer. High purity silver, and molybdenum targets were sputtered with Ar gas at 5 x 10$^{-3}$ mbar with RF and DC powers of than 25 W and 1 W to obtain deposition rates of 10 and 1.7 nm min$^{-1}$ respectively. Moreover, pure silicon target was also sputtered with a 10% O2/Ar gas mixture at 5 x 10$^{-3}$ mbar by applying a RF power of 100 W to form silicon oxide layers with a deposition rate of 7 nm min$^{-1}$. The selective solar coating, prepared using these growing conditions, is schematically represented in Fig. 1. From the bottom, the deposited stack is formed by four layers: (i) 250 nm thick silver layer acting as IR-mirror, (ii) 85 nm thick layer of high metal volume fraction (HMVF) cermet composed by Mo and SiO2 with 50% filling factor, (iii) 85 nm thick layer of low metal volume fraction (LMVF) cermet composed by Mo and SiO2 with 20% filling factor and, on top, (iv) 53 nm thick antireflective layer of SiO2.

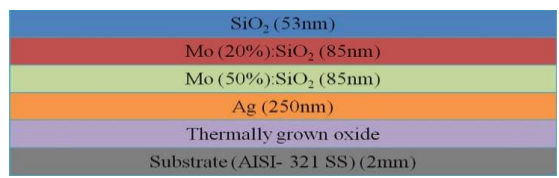

Fig. 1. Schematic representation of the selective solar coating used for the emissivity characterization.

Fig. 2 shows the measured reflectivity at the UV–vis–IR wavelength range to illustrate the selective character of such a multilayer coating. UV–vis–IR reflectivity measurements were performed using both a Shimadzu SolidSpec-3700 spectrophotometer, in the range of 0.19–3.30 µm, and a Varian 660-IR FTIR spectrometer in the 2.5–25 µm wavelength range. It can be easily observed the abrupt change in the reflectivity spectrum (R(λ)) from very low values at the UV–vis region to very high ones at the IR range, which makes possible to obtain a total solar absorptivity of α = 0.9, and a total thermal emissivity at room temperature of ε = 0.02. These values have been calculated by the well known expressions (1) and (2), using the measured

near normal reflectivity R(λ) in good approximation of the angle dependent R(λ,θ)

$$\alpha_s = \frac{\int_0^\infty [1-R(\lambda)]A(\lambda)d\lambda}{\int_0^\infty A(\lambda)d\lambda} \qquad (1)$$

$$\varepsilon_T(T) = \frac{\int_0^\infty [1-R(\lambda)]L(\lambda,T)d\lambda}{\int_0^\infty L(\lambda,T)d\lambda} \qquad (2)$$

where A(λ) is the Solar emission ASTM G173-03 Reference Spectrum (AM1.5) and L(T, λ) the Planck function.

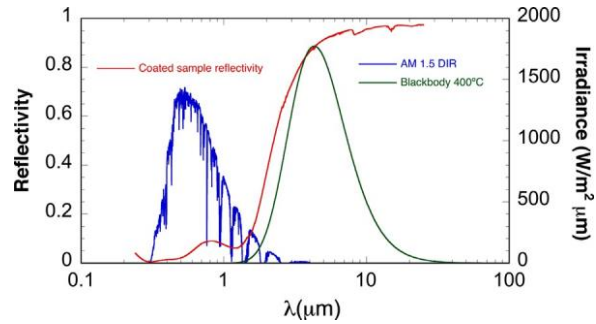

Fig. 2. UV–vis–IR reflectivity of the selective solar coating of Fig. 1.

The spectral emissivity measurements were carried out using a high accuracy infrared radiometer (HAIRL) described in Ref. [16], which allows accurate signal detection and fast processing. A diaphragm adjusts the sample area viewed by the detector and ensures good temperature homogeneity of the sample measured area. The sample holder permits directional measurements, while the sample chamber ensures a controlled atmosphere (vacuum, inert gas or open atmosphere). The set-up calibration was carried out by using a modified two-temperature method [17] and the emissivity was obtained applying the blacksur method [18]. The combined standard uncertainty of this direct emissivity device was previously obtained from the analysis of all uncertainty sources [19]. For the measurements presented in this paper, the maximum combined standard uncertainty varies between 1% and 9% depending on wavelength and temperature, its average value being around 3.5%. The sample temperature is measured by means of two K-type thermocouple spot-welded on the sample surface out of the area viewed by the detector. Before placing the sample in the sample holder its surface is cleaned in an ultrasonic bath of acetone. Once the sample is introduced in the sample chamber, the measurements were carried out in a moderate vacuum or with a slightly reducing atmosphere in order to prevent the oxidation of the sample surface. The measurements are performed during five heating cycles between room temperature and nearly 700 ºC. For each heating cycle the emissivity is measured at six or seven temperatures. It is interesting to note that in order to ensure the thermal equilibrium for each temperature, the time required for the measurements of a complete thermal cycle is one day.

In addition, direct emissivity measurements of the samples were obtained at 82 ºC using an emissometer model AE1 from Devices & Service Company.

3. Results and discussion

The following sections show the results and discussion of the total and spectral emissivity measurements obtained for the steel substrate (Section 3.1) and for the deposited selective coating (Section 3.2).

3.1. Total and spectral emissivity measurements of the steel substrate

The spectral emissivity $\varepsilon(\lambda, T)$ measurements were carried out on the steel substrate for five consecutive heating cycles in a slightly reducing atmosphere. As shown in Fig. 3 there are no significant differences in the values of the normal spectral emissivity between the first and fifth heating cycle for this steel. Therefore, these experimental results confirm that the substrate is free of possible surface tensions [20]. In addition, the substrate surface was analyzed after the five heating cycle by means of X-ray diffraction and no signs of oxidation were found.

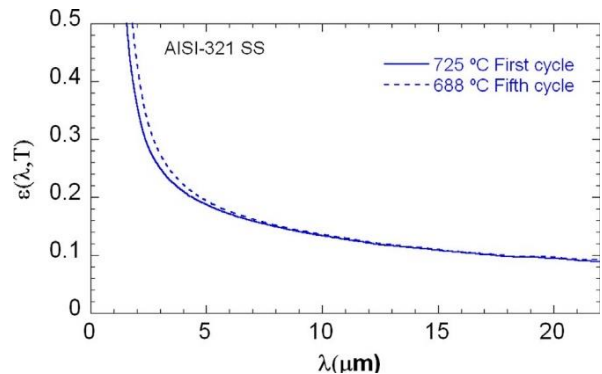

Fig. 3. Normal spectral emissivity $\varepsilon(\lambda, T)$ for substrate at 700 ºC as a function of wavelength for the first and fifth heating cycles.

Fig. 4 shows the normal spectral emissivity for seven temperatures, between 173 and 688 ºC, during the fifth heating cycle. According the electromagnetic theory, the emissivity decreases as wavelength increases [21]. However, it can be observed that the behavior of the emissivity with temperature undergoes a change around $\lambda = 3$ μm. This is the so-called X point for the AISI-321 SS sample. Above the X point the emissivity shows a slight temperature increase with an almost linear dependence, whereas for $\lambda \leq 3$ μm the emissivity decreases with temperature until $T = 350$ ºC. Experimental results in Fig. 4 show that for $\lambda = 10$ μm emissivity increases a 20% between 173 and 688 ºC. This result should be taken into account in the final design of the coated steel system.

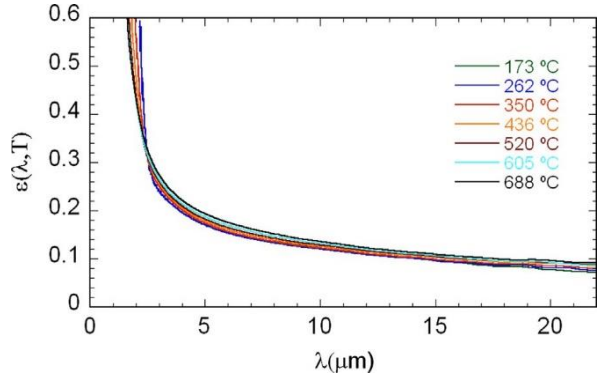

Fig. 4. Substrate normal spectral emissivity ε(λ, T) for the fifth heating cycle at different temperatures.

The temperature dependence of the total normal emissivity $\varepsilon_T$ (T), given by Eq. (3), is needed in order to calculate the radiation energy losses of a solar collector. However, calculations of this parameter have not been made at the typical working temperatures of solar collectors (T≥500 ºC).

$$\varepsilon_T(T) = \frac{\int_0^\infty \varepsilon(\lambda, T) L(\lambda, T) d\lambda}{\int_0^\infty L(\lambda, T) d\lambda} \quad (3)$$

The difference with Eq. (1) is the measured temperature dependence for the spectral emissivity. The integration of Eq. (3) requires to know the emissivity for wavelengths between 0 and 1, but we have experimental data between λ = 1.42 mm and λ = 22 mm. In this case the integral can be calculated from the following equation:

$$\int_0^\infty \varepsilon(\lambda, T) L(\lambda, T) d\lambda = \int_0^{\lambda_1} \varepsilon_1(\lambda, T) L(\lambda, T) d\lambda + \int_{\lambda_1}^{\lambda_2} \varepsilon(\lambda, T) L(\lambda, T) d\lambda$$
$$+ \int_{\lambda_2}^\infty \varepsilon_2(\lambda, T) L(\lambda, T) d\lambda \quad (4)$$

Thus, the total normal emissivity depends on the values of the spectral emissivity used as $\varepsilon_1(\lambda,T)$ and $\varepsilon_2(\lambda,T)$ in Eq. (4). To estimate the maximum range of variation of the total normal emissivity two extreme cases have been taken into account. For the highest value, $\varepsilon_1(\lambda,T) = 1$ is taken and for $\varepsilon_2(\lambda,T)$ the value of the normal spectral emissivity at $\lambda_2$ is extrapolated. The lowest value is obtained with $\varepsilon_2(\lambda, T) = 0$ and taking as $\varepsilon_1(\lambda, T)$ the value of the normal spectral emissivity at $\lambda_1$. The values for the two limits of the total normal emissivity for T = 655 ºC are 0.2186 and 0.2163 respectively, and the average value is 0.217±0.001. In Fig. 5 the total normal emissivity is plotted as a function of temperature. As it can be noted, it shows the linear behavior predicted by the electro- magnetic theory for metals.

Since the first layer of the selective coating is a steel thermally oxidized, an in situ thermal oxidation of the substrate was carried out in order to study the variation of the steel emissivity with the oxide thickness. Fig. 6 shows the normal spectral emissivity during the oxidation process of the substrate at 600 ºC for oxidation times up to 22 h. As expected, the emissivity increase with the oxidation time. In addition, it can be observed the first interferential maximum and minimum associated to the growth of the oxide layer for t > 10 h [22]. The emissivity behavior of this steel is similar to other metals [22].

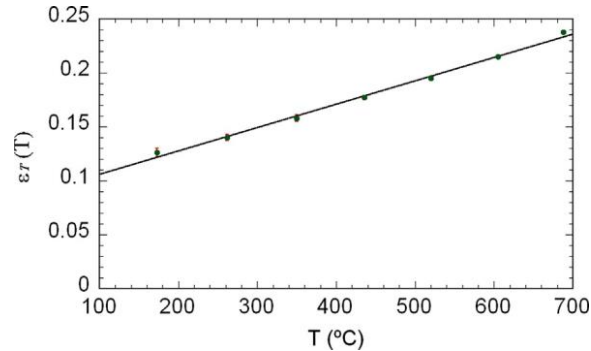

Fig. 5. Total normal emissivity $\varepsilon_T(T)$ of substrate as a function of temperature.

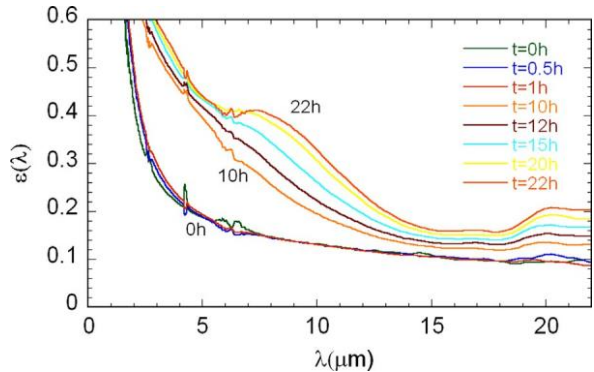

Fig. 6. Spectral normal emissivity $\varepsilon(\lambda)$ of the substrate for various times during the oxidation process in air at 600 ºC. The first interferential maxima and minima can be observed.

*3.2.* Spectral emissivity measurements of the selective solar coating

The emissivity measurements of the coating were carried out between 150 and 600 ºC and under moderate vacuum (~$10^{-3}$ mbar), according to the working conditions in the solar collectors. The first step has been the study of the emissivity during heating with a maximum heating rate of 2 ºC min$^{-1}$. In the measurement method used in this study the temperature is stabilized during 20 min after every 30 ºC step approximately, in order to measure the emissivity. Fig. 7 shows the emissivity values as a function of the heating cycle for two temperatures and four wavelengths. It can be observed that the emissivity variations between the first and fourth heating cycle are lower than the experimental uncertainty. It was then checked that, at this heating rate, the coating remains unchanged. The results for all wave- length and temperature ranges suggest that the coating is stable over the life of the solar collector. In order to verify this statement, further tests to simulate one year operation are underway, using a programmed heating system.

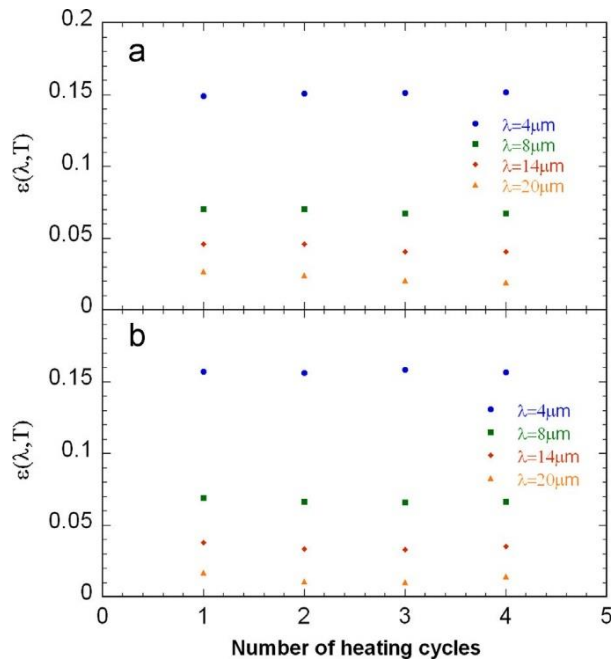

Fig. 7. Coating emissivity ε(λ, T) as a function of the heating cycle for 4, 8, 14 and 20 µm wavelengths for (a) T = 320 ºC and (b) T = 600 ºC.

Since the operative heating rate in a solar collector is around 10 ºC min$^{-1}$, the emissivity was measured as a function of the heating rate in a range between 2 and 10 ºC min$^{-1}$ in order to analyze the stability of the optical properties of the SASs The results showed emissivity variations smaller than the experimental uncertainty. Therefore, it can be stated that the coating emissivity is independent of the heating rate, which is a critical feature of the coating in order to its application in a real solar plant.

Fig. 8 shows the normal spectral emissivity for eight temperatures between 236 and 600 ºC during the fourth heating cycle. It can be noticed that these emissivity spectra show the same behavior of the bare steel substrate, with lower emissivity values (Fig. 4). In this case the point X is also observed around 3 µm. From the comparison of the normal spectral emissivity of the substrate and coating (Figs. 4 and 8) it is concluded that the emissivity of the coating is significantly lower than that of the substrate, in the most important range for thermal radiation (λ > 2 µm), between 10% and 30% for low and high wavelength respectively. In addition, one can state that this coating presents the optimal emissivity values required in high temperature solar collectors, ε < 0.05 for λ > 10 µm.

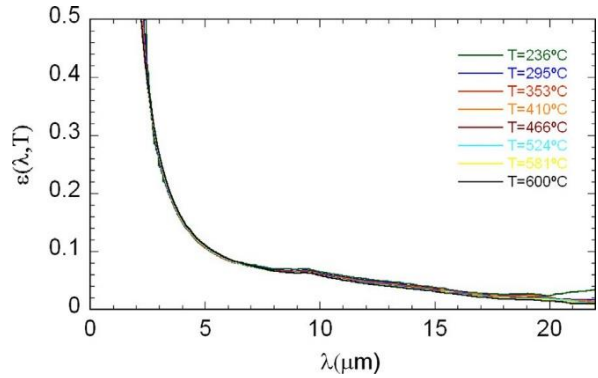

Fig. 8. Normal spectral emissivity ε(λ, T) of selective coating for eight different temperatures measured in the fourth heating cycle.

Fig. 9 represents the emissivity as a function of temperature for four different wavelengths. The emissivity shows an almost negligible decrease with temperature except for small wavelengths. For this spectral range, the emissivity decreases slightly until 450 ºC and remains constant up to 600 ºC, showing a good performance for high temperature solar harvesting.

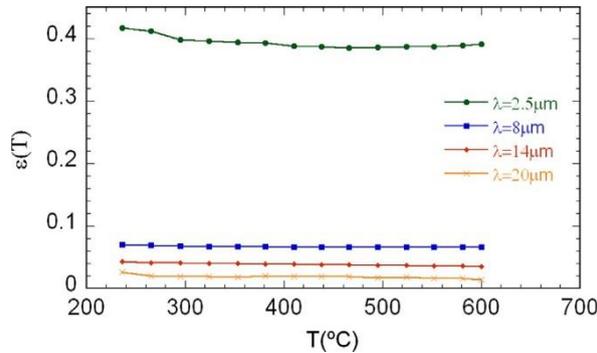

Fig. 9. Normal spectral emissivity ε(T) of the selective coating measured at the fourth heating cycle as a function of temperature for four different wavelengths.

In order to calculate the radiation energy losses of a solar collector, it is necessary to know the total normal emissivity. Fig. 10 shows the total normal emissivity obtained for the coated SAS sample using three different experimental methods. In first place, the emissivity value obtained with a commercial emissometer operating at 82 ºC is included (black star in Fig. 10). Secondly, the emissivity values obtained from the reflectance spectrum measured at room temperature (full triangles in Fig. 10) are plotted. In this case, the typical procedure to estimate emissivity at high temperatures using Eq. (2) has been applied, where the temperature dependence comes from the Planck function and the reflectance spectra, R(λ), is considered temperature-independent. In agreement with the discussion of Eq. (4) in Section 3.1, the integration limits can be taken from 1 to 30 μm. Furthermore, it must be mentioned that the reflectometer reference produces small errors in the absolute reflectance values. To avoid this uncertainty, the reflectance spectrum in the 1.5–30 μm range has been shifted to coincide with the emissometer value at 82 ºC. Finally, a rigorous calculation of the total emissivity was carried out for the first time using Eqs. (3) and (4) (close circles in Fig. 10). In this case, both Planck function and experimental normal spectral emissivity are temperature dependent. In the same figure, the total normal emissivity of the substrate is plotted for comparison (full squares).

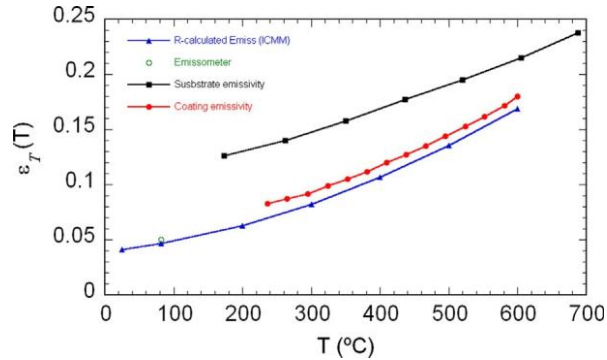

Fig. 10. Total normal emissivity $\varepsilon_T(T)$ of SAS obtained from radiometric measurements (full circles), from reflectivity measurements (full triangles) and with an emissometer at 82 ºC (open circle). The substrate results are also plotted (full squares).

These results shown in Fig. 10 allow an interesting analysis of this type of selective coatings as well as a comparison between the experimental measurements methods used in this paper. First, it is important to note that the temperature dependence of the total normal emissivity obtained from the reflectivity at room temperature shows the same qualitative behavior of the obtained using the normal spectral emissivity, which is temperature dependent (Fig. 10). Only small quantitative differences are appreciated in Fig. 10. The close agreement between radiometric and reflectance measurements in Fig. 10 is caused by the very weak temperature dependence of the spectral emissivity of the SAS (see Fig. 9). Just this point justifies the interest of measuring the temperature dependence of the normal and, in some cases, also the directional spectral emissivity. In addition, it is essential, for solar selective coatings applications to assess the emissivity changes with heating cycles and heating rate, which can only be proven with spectral measurements at different temperatures. Another important aspect for the applications that can be studied by means of radiometric methods is the detection of anomalous behavior of the SAS with the temperature and wavelength. Finally, if a coating has a spectral emissivity with a significant decrease with temperature, the total emissivity calculated from Eq. (3) will be lower than those calculated from Eq. (2) on the basis of the same value at room temperature. This is because the major temperature dependence of the emissivity obtained from Eq. (2) arises from the Planck function. Thus, all the curves obtained from this equation will have similar temperature behavior modulated by the shape of the reflectance spectrum.

The results of Fig. 10 also suggest some comments about the coating itself. Firstly, the total normal emissivity of coated sample shows a slight curvature with the temperature while the substrate has a quasi-linear behavior according to electromagnetic theory predictions. Secondly, the difference between the experimental values of the total normal emissivity between the substrate and the coating shows the effectiveness of the latter and indicates that it has an appropriate thickness. Finally, it is important to note that the results in Fig. 9 together with the design flexibility and thermal stability of the

SAS open up the possibility to compensate the Planck function shift with temperature. Experimental work in this direction is underway in our laboratory.

4. Conclusions

In this paper, it is presented, for the first time, a complete radiometric characterization of a selective absorber surface by using absolute measurements of spectral emissivity at the whole working temperature range (150–600 ºC). The total emissivity values obtained from spectral emissivity measurements are com- pared with those obtained from reflectivity data and with direct characterization by a commercial emissometer at 82 ºC. This spectral emissivity method allows to know the actual behavior of coated stainless steel system at the operation temperatures for CSP applications as it is the unique test to detect any anomalous behavior of the coating with temperature and wavelength. Only in the case where the spectral emissivity does not vary with temperature within the wavelength range in which the radiative transfer is made, a qualitative agreement between reflectivity measurements and the radiometer could be acceptable.

As expected, for a selective coating, the results show that coated stainless steel present a considerable lower emissivity when compared with bare stainless steel. It is worth noting, the assessment of the emissivity stability with temperature and the thermal cycling under different heating rates, which ensures the performance for the required application.

Acknowledgements

This work was financially supported by the European Commission (project HITECO FP7-ENERGY-2010-ºCollaborative N. 256830) the Spanish Ministry of Science and Innovation (projects FUNCOAT CSD2008-00023, RyC2007-0026) and program ETORTEK of the Consejería de Industria of the Gobierno Vasco in collaboration with the CIC-Energigune Research Center. L. González-Fernández acknowledges the Basque Government the support through a Ph.D. fellowship.

References

[1] C.E. Kennedy. Review of Mid-to-High-Temperature Solar Selective Absorber Materials. NREL/TP-520-31267. Available from: ⟨http://www.nrel.gov/docs/fy02osti/31267.pdf⟩.

[2] G.A. Niklasson, C.G. Granqvist, Selective solar-absorbing surface coatings: optical properties and degradation, in: C.G. Granqvist (Ed.), Materials Science for Solar Energy Conversion SystemsPergamon Press, Oxford, 1991, p. 70.

[3] M. Nejati. Cermet based solar selective absorbers; further selectivity improvement and developing new fabrication technique, Ph.D.-Dissertation, Saarbrücken, 2008.

[4] O.T. Inal, A. Sherer, Optimization and microestructural analysis of electro- chemically deposited selective solar absorber coatings, Journal of Materials Science 21 (1986) 729–736.

[5]   L. Katulza, A. Surca-Vuc, B. Orel, Structural and IR spectroscopy analysis of sol–gel processed CuFeMnO4 spinel and CuFeMnO4/silica films for solar absorbers, Journal of Sol–Gel Science and Technology 20 (2001) 61–83.

[6]   Z.C. Orel, M.K.Gunde, Spectral selective paint coatings preparation and characterization, Solar Energy Materials and Solar Cells 68 (2001) 337–353.

[7]   M. Farooq, M.G. Hutchins, A novel design in composites of various materials in solar selective coatings, Solar Energy Materials and Solar Cells 71 (2002) 523–535.

[8]   I.T. Ritchie, B. Window, Application of thin graded index films to solar absorbers, Applied Optics 16 (1977) 1438–1443.

[9]   M.R. Nejati, V. Fathollahi, M.K. Asadi, Computer simulation of the optical properties of high temperature cermet solar selective coatings, Solar Energy 78 (2005) 235–241.

[10]  B.O. Seraphin, Thin films in phototermal solar energy conversion, Thin Solid Films 90 (1982) 395–403.

[11]  N. Selvakumar, H.C Barshilia, Review of physical vapor deposited (PVD) spectrally selective coatings for mid- and high-temperature solar thermal applications, Solar Energy Materials and Solar Cells 98 (2012) 1–23.

[12]  Q.-C. Zhang, Recent progress in high-temperature solar selective coatings, Solar Energy Materials and Solar Cells 62 (2000) 63–74.

[13]  S. Esposito, A. Antonaia, M.L. Addonizio, S. Aprea, Fabrication and optimisation of highly efficient cermet-based spectrally selective coatings for high operat- ing temperature, Thin Solid Films 517 (2009) 6000–6006.

[14]  C.E. Kennedy, H. Price, Progress in development of high-temperature solar-selective coatings, NREL/CP-520-36997. in: Proceedings of the 2005 Interna- tional Solar Energy Conference, ISEC'05, Orlando, Florida, USA, ISEC2005- 76039, August 6–12, 2005.

[15]  C.G. Granqvist, Transparent conductors as solar energy materials: a panoramic review, Solar Energy Materials and Solar Cells 91 (2007) 1529–1598.

[16]  L. del Campo, R.B. Pérez-Sáez, X. Esquisabel, I. Fernández, M.J. Tello, New experimental device for infrared spectral direction emissivity measurements in a controlled environment, Review of Scientific Instruments 77 (2006) 113111.

[17]  L. González-Fernández, R.B. Pérez-Sáez, L. del Campo, M.J. Tello, Analysis of

calibration methods for direct emissivity measurements, Applied Optics 49 (2010) 2728–2735.

[18]  R.B. Pérez Sáez, L. del Campo, M.J. Tello, Analysis of the accuracy of methods

for the direct measurement of emissivity, International Journal of Thermo- physics 29 (2008) 1141–1155.

[19] L. del Campo, R.B. Pérez-Sáez, L. González-Fernández, M.J. Tello, Combined standard uncertainty in direct emissivity measurements, Journal of Applied Physics 107 (2010) 113510.

[20] L. Gonzalez-Fernández, E. Risueño, R.B. Pérez Sáez, M.J. Tello, Infrared normal spectral emissivity of Ti–6Al–4V alloy in the 500–1150 K temperature range, Journal of Alloys and Compounds 541 (2012) 144–149.

[21] M.F. Modest, Radiative Heat Transfer, 2nd ed., Academia Press, California, 2003, pp. 76-84.

[22] L. del Campo, R.B. Pérez-Sáez, M.J. Tello, X. Esquisabel, I. Fernández, Armco iron normal spectral emissivity measurements, International Journal of Thermophysics 27 (2006) 1160–1172.